\documentclass[boxit]{JAC2003}

\usepackage{graphicx}
\usepackage{booktabs}
\usepackage{verbatim}

\begin{document}
\title{Project PROMETHEUS: \\Design and Construction of a Radio Frequency Quadrupole at TAEK}

\author{G. Turemen\thanks{gorkem.turemen@ankara.edu.tr}, B. Yasatekin\\ \emph{Ankara University, Department of Physics, Ankara, Turkey.}\\ O. Mete\\ \emph{Cockcroft Institute, Warrington and University of Manchester, UK.}\\ M. Celik, Z. Sali\\ \emph{Gazi University, Department of Physics, Ankara, Turkey.}\\ Y. Akgun, A. Alacakir, S. Bolukdemir, E. Durukan, H. Karadeniz, E. Recepoglu\\ \emph{TAEK, SANAEM, Ankara, Turkey.}\\  E. Cavlan\\ \emph{TOBB ETU, Department of Electrical and Electronics Engineering, Ankara, Turkey.}\\ G. Unel\\ \emph{University of California at Irvine, Physics Department, Irvine, California, USA.} \\S. Erhan\\ \emph{University of California at Los Angeles, Department of Physics and Astronomy, Los Angeles, California, USA.}}

\maketitle
\begin{abstract}
The PROMETHEUS Project is ongoing for the design and development of
a 4-vane radio frequency quadrupole (RFQ) together with its $H^+$ ion source,
a low energy beam transport (LEBT) line and diagnostics section. The
main goal of the project is to achieve the acceleration of the low
energy ions up to 1.5 MeV by an RFQ (352 MHz) shorter than 2 meter. A
plasma ion source is being developed to produce a 20 keV, 1 mA $H^+$
beam. Simulation results for ion source, transmission and beam dynamics are presented together with
analytical studies performed with newly developed RFQ design code DEMIRCI. Simulation results shows that a beam transmission 99\% could be achieved
at 1.7 m downstream reaching an energy of 1.5 MeV. As the first phase
an Aluminum RFQ prototype, the so-called cold model, will be built
for low power RF characterization. In this contribution the status
of the project, design considerations, simulation results, the various
diagnostics techniques and RFQ manufacturing issues are discussed.
\end{abstract}

\section{Introduction}

An energetic proton beam is indeed a very useful tool: It's usage ranges
from basic scientific research (like in the case of the LHC) to a
medical application (such as Hadron Therapy) or from production of
secondary beamlines to simply being a tool for educational and test purposes.
The common properties of all proton (linear) beamlines is to start
with an ion source, followed by a Low Energy Beam Transport (LEBT),
and an accelerating structure effective at low beta ($\beta=v/c$). The structure
of choice for the accelerating structure is a Radio Frequency Quadrupole (RFQ) since
1980s. The PROMETHEUS project, at the Turkish Atomic Energy Authority's
(TAEK) Saraykoy Nuclear Research and Training Center (SANAEM), aims
to gain the necessary knowledge and experience to construct such a proton beam.
A Proof of Principle (PoP) accelerator with modest requirements of achieving at least 1.5
MeV proton energy, with an average beam current of at least 1 mA is under development. PoP project also have the challenging goal of having the design and construction of the
entire machine in Turkey, from its ion source up to last diagnostic
station, including its RF power supply and to complete it in three
years. There are also two secondary goals of this project: 1) Training accelerator physicists and RF engineers on
the job; 2) To encourage local industry in accelerator component
construction.

\section{Beamline Design}

\begin{figure}[htb]
   \centering
   \includegraphics*[width=80mm]{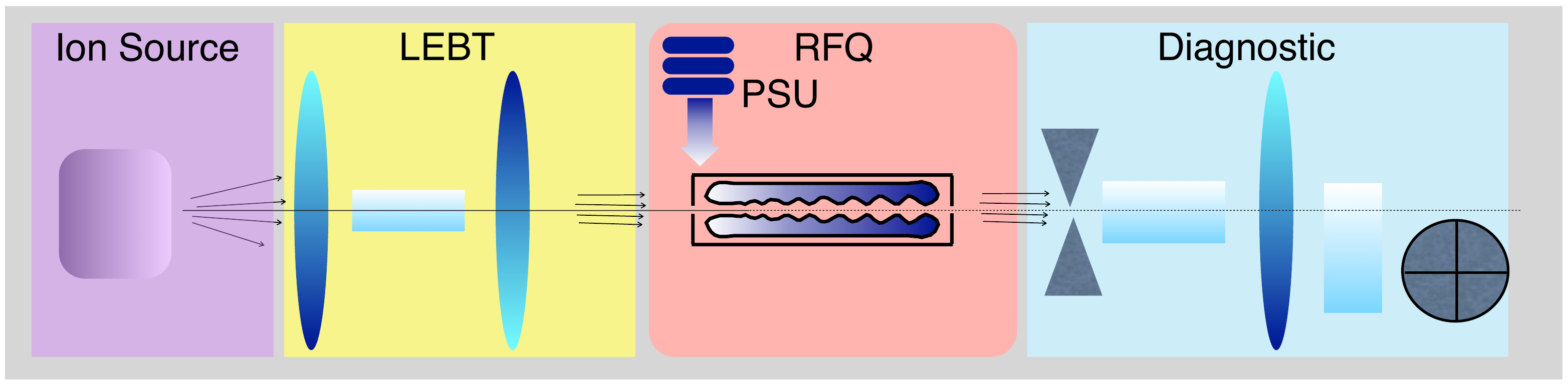}
   \caption{Layout of the SANAEM Beamline.}
   \label{fig:The-planned-beamline}
\end{figure}

The beamline design originates from previous setups and past experience of SANAEM.
It consists of an ion source, a LEBT,
a RFQ and a beam dump with diagnostic
stations inserted at the appropriate places. A schematic view of the
beamline can be seen in Fig. \ref{fig:The-planned-beamline}.

\subsection{Ion Source}

\begin{figure}[htb]
   \centering
   \includegraphics*[width=80mm]{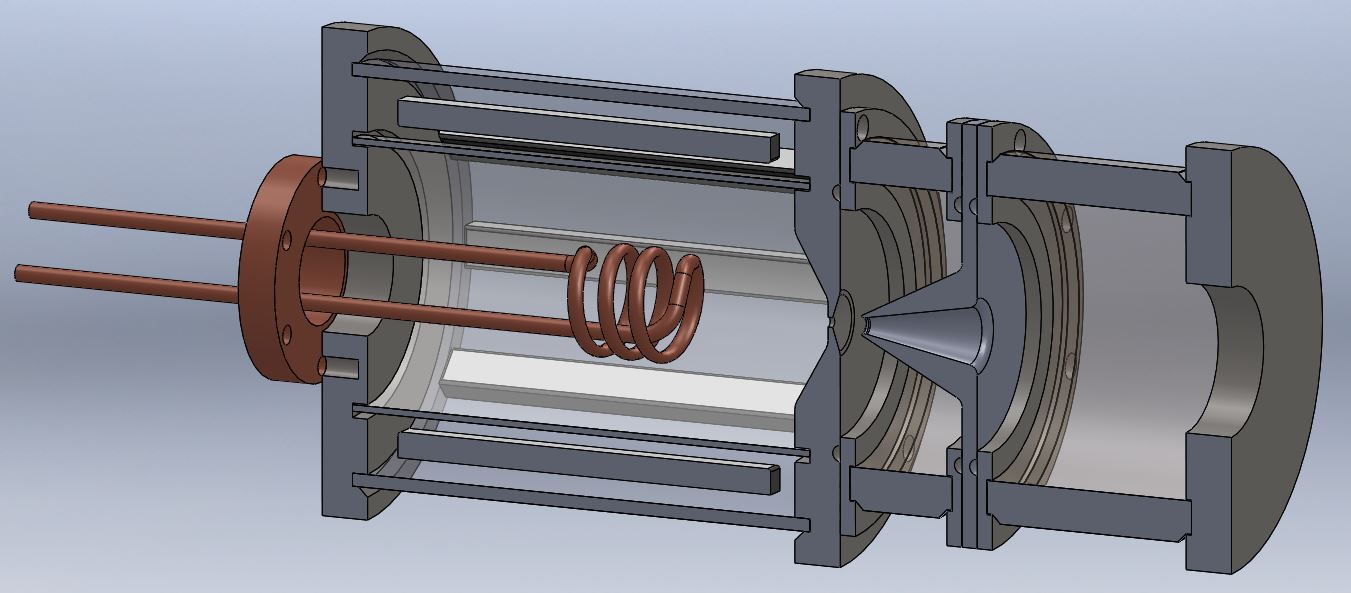}
   \caption{3D Drawing of the Ion Source.}
   \label{is}
\end{figure}

The $H^+$ ion source, consisting of two parts (plasma generator and extraction
system) is being redesigned to match the required input parameters of the main
accelerator structure, i.e. the RFQ. It should give as much current as
possible at a low energy to keep the RFQ length as short as possible.
Therefore, the main parameters of the ion source are: 20 keV output
energy, at least 1mA of average beam current and a total normalized
transverse emittance smaller than 1.5 $\pi$.mm.mrad. The actual design
is realized with IBSimu\cite{ibsimu} and SimIon\cite{simion} software
packages. Although the ion plasma can be generated by various methods,
this work focuses on two options: By heating a filament with DC to
provoke thermionic emission and alternatively by using an RF antenna.
The plasma confinement system is achieved with 8 rod permanent multicusp
magnets with a field of 4000 Gauss each. The requirement of using $H^+$ ions
simplifies the extraction system to only 3 electrodes. The whole system
is designed to be water-cooled housed in a Plexiglas vessel. The 3D representation of the ion source currently under construction is shown in Fig \ref{is}.

\subsection{LEBT}

The LEBT is used to transport the beam exiting the ion source with a minimum
loss of particles up to the input of the RFQ and also to match its
input parameters. The desired beam parameters will be achieved by
the use of two solenoid magnets of length 15 cm each. The LEBT design
is obtained with TRAVEL \cite{travel} software library and the solenoids
themselves by the use of SUPERFISH \cite{superfish}. The magnetic
fields of the two solenoids have been calculated as 2830 Gauss at $z=7.5$
cm and 2480 Gauss at $z=72.5$ cm.

\subsection{RFQ}

The decision on the ion type, namely the selection of $H^+$, also dictates
the type of the RFQ as 4-vane type. Considering the local manufacturing capabilities and constraints, the inter-vane voltage was selected
to be 60 kV, constant throughout the RFQ. Similar arguments lead to
1.5 as the Kilpatrick value. The decision on the RF operation frequency was not based on the readily available material but
on the collaboration possibilities with CERN. Choice of 352.21 MHz operations frequency is compatible with the 4-vane decision and also results in a compact size RFQ cell that is compatible with the local CNC (Computer Numerical Control) capabilities. The
actual design is made by the use of two computer programs (LIDOS \cite{lidos} and DEMIRCI \cite{dmrc} discussed below.), cross-checking
each other.

The SANAEM RFQ has been designed using LIDOS, one of the two commercially
available computer programs. The main design parameters can be found in Table \ref{par}. The preliminary result from "LIDOS.Advisor",
satisfies the initial design requirements, namely obtaining 1.5 MeV
energy in less than 2 meters with two sections less than 90 cm each.
This software package provides a total solution including the final
vane design files that can be fed directly to the CNC machines. However some difficulties in running it in batch mode and in obtaining some design parameters as a text file lead us to develop a similar computer program, DEMIRCI, with local resources.

\begin{table}[hbt]
   \centering
   \caption{RFQ Main Design Parameters.}
   \begin{tabular}{ccc}
       \toprule
        Parameter & Value & Unit  \\
       \midrule
        
          $E_{in}$ & 20 & keV    \\
           $E_{out}$ & 1.5 & MeV    \\
          $f_{cav}$ & 352.21 & MHz     \\
	I & 1 & mA \\
$\epsilon_N$ & 1 & $\pi$.mm.mrad \\
$V_o$ & 60 & kV \\
KP & 1.5 & - \\
$R_{0}$ & 2.799 & mm \\
$\rho$ & 2.5 & mm \\
       \bottomrule
   \end{tabular}
   \label{par}
\end{table}

\begin{figure}[hbt]
   \centering
   \includegraphics*[width=75mm]{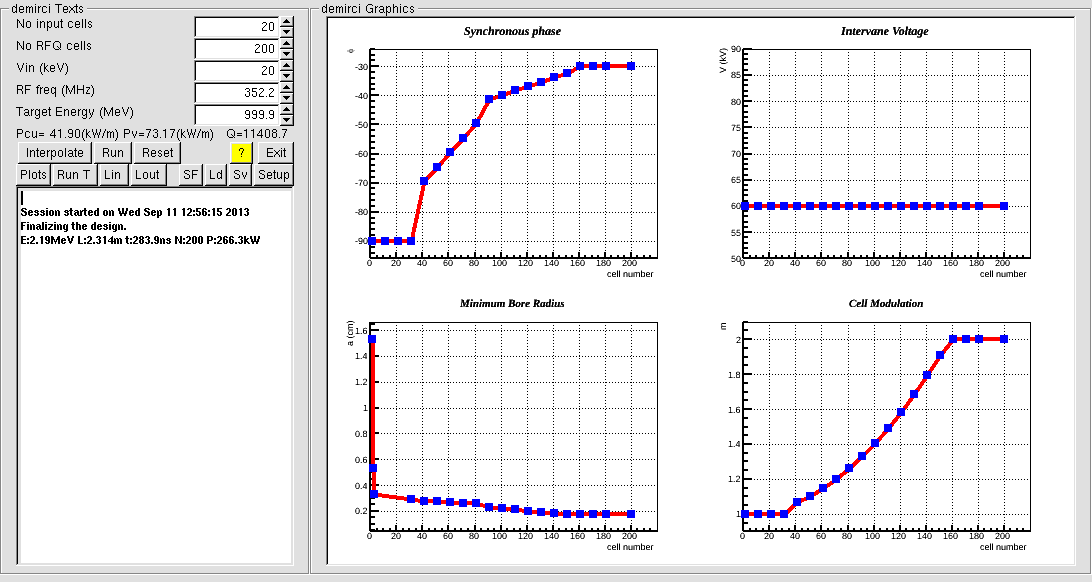}
   \caption{A Snapshot from DEMIRCI's GUI.}
   \label{dmr}
\end{figure}

DEMIRCI, uses the classical RFQ
formulas based on two term potential. It incorporates ROOT \cite{root} library
for user interaction and plotting facilities (see Fig. \ref{dmr}). It interfaces other
available software like LIDOS, TOUTATIS \cite{toutatis} and SUPERFISH. It has also
a command line interface, designed to be used in batch mode, where
GNUPLOT is used for plotting. Although DEMIRCI results are found to
be compatible with LIDOS and TOUTATIS within few percent error margin (see Table \ref{comp}), its development
continues as it does not yet provide CNC compatible vane design files. The difference between the results is believed to be originated from potential function expansions, i.e. 8 term multipole versus classical 2 term potential.

\begin{table}[hbt]
   \centering
   \caption{DEMIRCI-TOUTATIS-LIDOS Comparisons.}
   \begin{tabular}{cccc}
       \toprule
        Parameter & DEMIRCI & LIDOS & TOUTATIS  \\
       \midrule
       
          Length (m) & 1.555 & 1.585 & 1.549    \\
           Energy (MeV) & 1.54 & 1.52 & 1.49    \\
           Time (ns) & 249.9 & 265.8 & 243.8     \\

              Transmission(\%) & n/a & 99 & 97 \\

       \bottomrule
   \end{tabular}
   \label{comp}
\end{table}

\section{RFQ Beam Dynamics}

\begin{figure}[htb]
   \centering
   \includegraphics*[width=80mm]{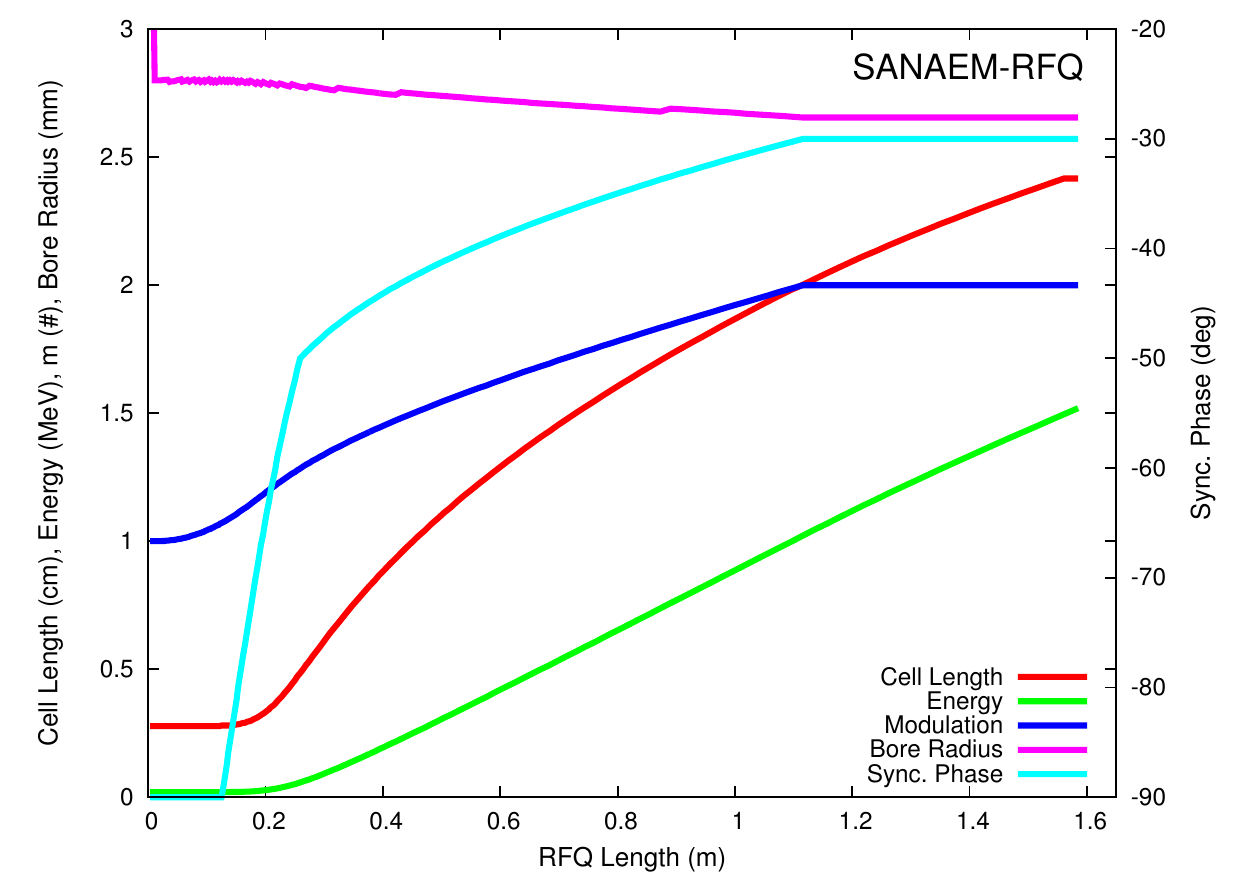}
   \caption{Beam Dynamics Parameters Along the RFQ. }
   \label{lanl}
\end{figure}

\begin{figure}[htb]
   \centering
   \includegraphics*[width=85mm]{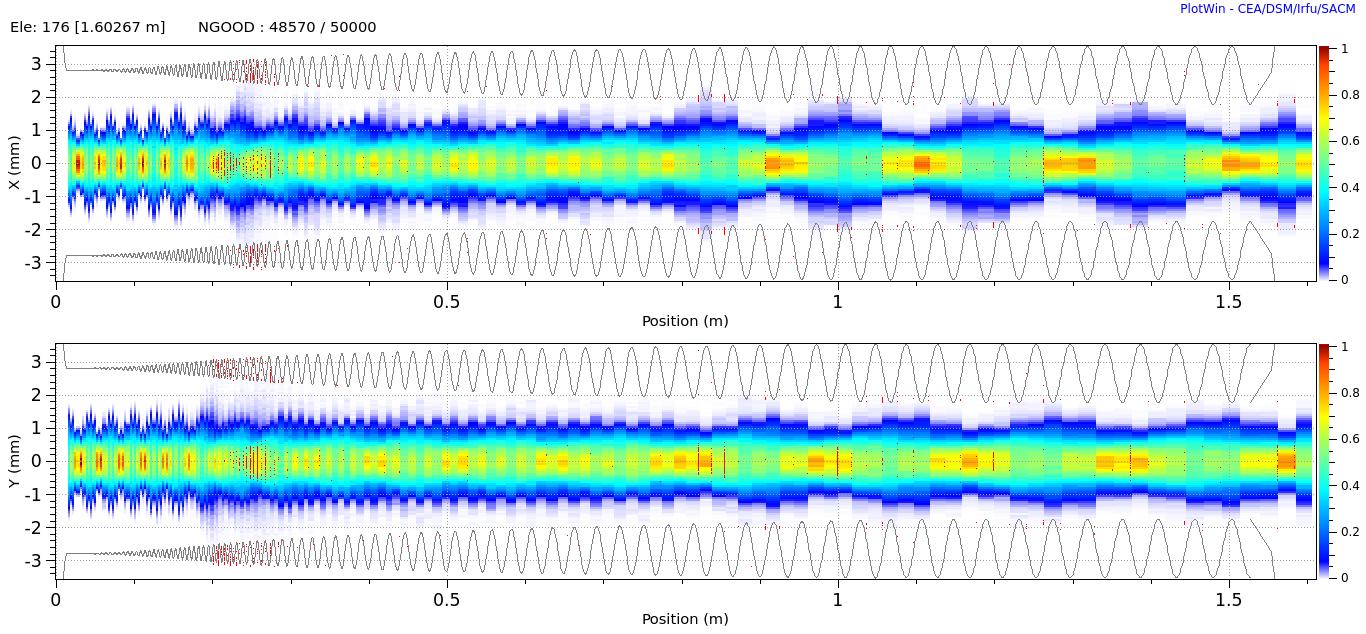}
   \caption{Beam Profiles Along the RFQ.}
   \label{env}
\end{figure}

The SANAEM RFQ was designed using the Los Alamos National Laboratory's (LANL) traditional Four Section
Procedure (FSP). After the static design, the real vane shapes and actual beam dynamics
were primarily calculated by "LIDOS.RFQ.Designer" simulation code. In addition, a similar computer program, TOUTATIS, was used for beam dynamics design to validate the preliminary design by comparing it's results with LIDOS. Two beam dynamics results obtained with TOUTATIS and LIDOS were compared with each other with the use of relevant DEMIRCI module. Additionally, LIDOS
enables the segmentation of the RFQ design to the be with CNC
manufacturing capabilities. The dividing gap is selected to be at $z=80$ cm with
a separation of 1 mm. The resulting total structure is about 156 cm long,
consisting of 176 cells with an inter-vane distance of 2.8 mm and a
constant vane tip radius of 2.5 mm. The mismatch ratio is given by
LIDOS as 1.03, which is a measure of the emittance match between the
beam at the end of the LEBT and the ideal beam accepted by the RFQ. Also one should note that as the average current is modest,
the space charge effects are negligible. 
Finally, the RFQ transmission is found to be about 99\% for all particles
and about 96\% for accelerated particles. The beam dynamics design
results are shown in Fig. \ref{lanl} and Fig. \ref{env}. 

SANAEM RFQ exit ends with a Crandall cell which makes the beam properties
time independent by canceling the phase difference between x and y
dimensions. The proton beam at the RFQ exit can be guided by solenoid
magnets into the next structure for further acceleration.

\section{RFQ Electromagnetic Considerations}

\begin{figure}[htb]
   \centering
   \includegraphics*[width=80mm]{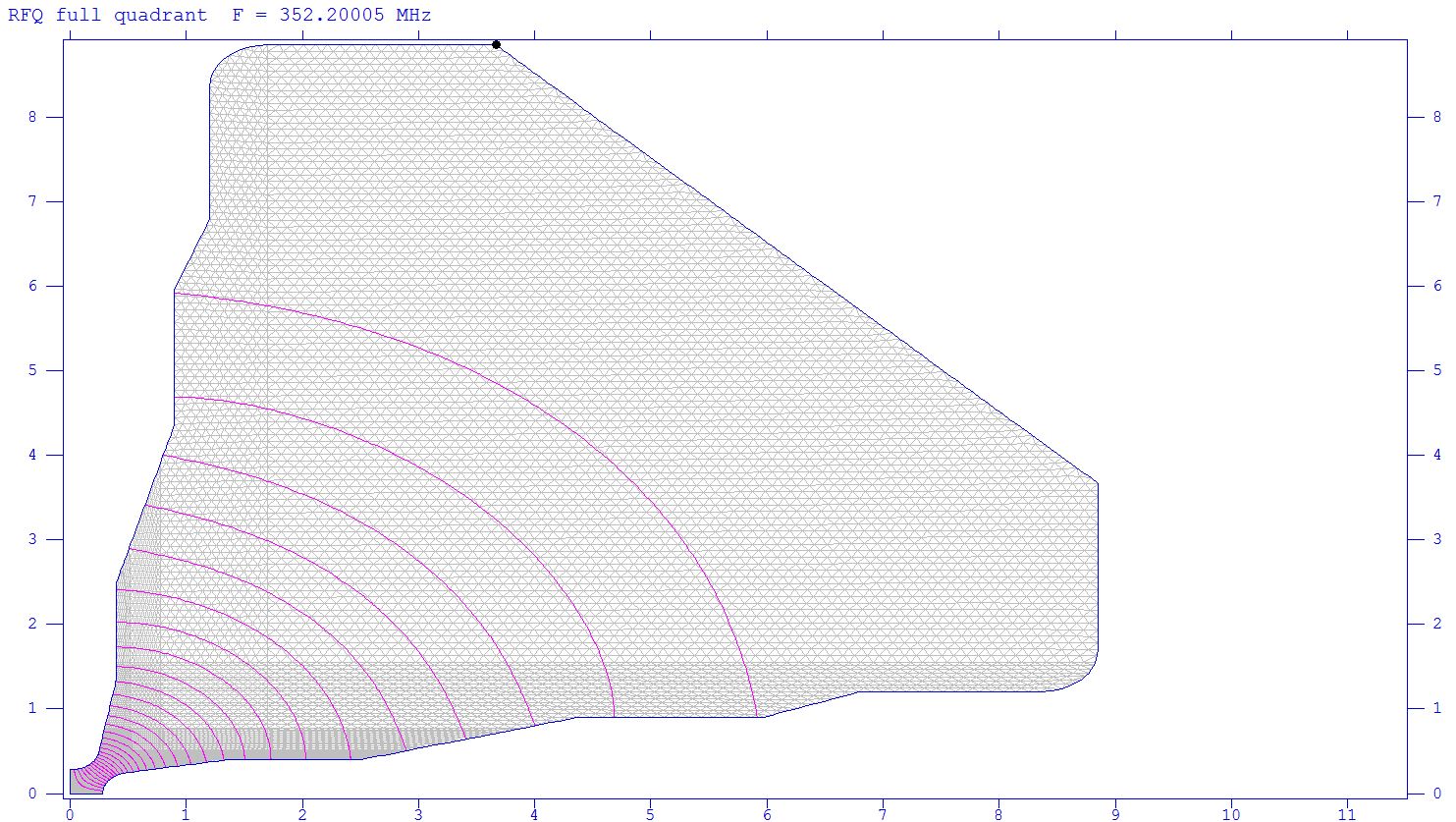}
   \caption{2D Electromagnetic Design of the SANAEM RFQ.}
   \label{2d-design}
\end{figure}

SANAEM RFQ electromagnetic studies were performed by using SUPERFISH,
RFQFish and CST MWS \cite{cst} simulation programs to fine tune the cavity physical
parameters. The first step was to start with the beam dynamics parameters
and to achieve a preliminary design in 2D with Superfish which was
subsequently optimized with RFQfish. The vane structure obtained after
this last step was an input to CST MWS program which was used for
3 dimensional design. The cross sectional view and the geometrical
properties can be seen in Fig. \ref{2d-design}.

\begin{figure}[htb]
   \centering
   \includegraphics*[width=75mm]{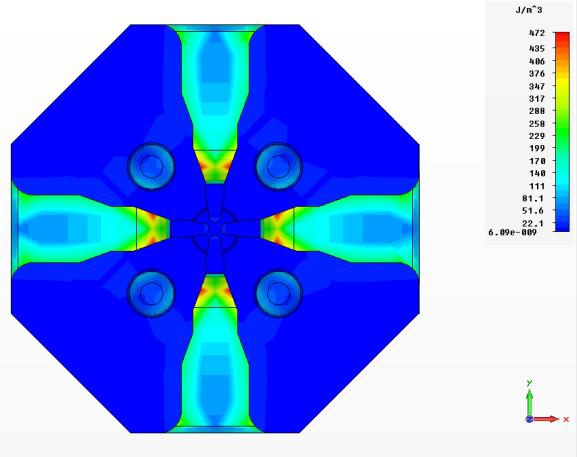}
   \caption{Electrical Energy Density of the SANAEM RFQ in 3D Cross-section. }
   \label{3dcross}
\end{figure}

The quadrupole and dipole modes of the SANAEM RFQ are calculated
by applying Neumann and Dirichlet boundary conditions and found 352.199 MHz with a quality factor of 10342 for the former, whereas
the latter is at 341.903 MHz with a quality factor of 10173. The electrical energy density inside the SANAEM RFQ is shown in Fig \ref{3dcross}. The power loss on
the RFQ walls per quadrant is calculated to be 126 W/cm which gives
a total ohmic loss of 91 kW at a total length of 164 cm which also include input and output matching sections. The volume
power requirement has been calculated as 115 kW, which gives total power
requirement of 206 kW. Including a conventional safety margin of 25\% ($\sim$50 kW) to account
for losses in the transmission lines and RF connection, the RFQ power
supply has to deliver 250 kW. Following the project principles, the
RF power supply will be built by the local industry under the guidance
of the project members. The RFQ operation will start in pulse mode
with a relatively low duty factor of 5\%. During the later stages of
the operation, this value will be increased by upgrading the RF psu. The relevant
(brief) list of physical parameters for the cavity is given in Table \ref{reff}.

\begin{table}[hbt]
   \centering
   \caption{RFQ Cavity Physical Parameters After 3D Design\label{reff}}
   \begin{tabular}{ccc}
       \toprule
       Parameters & Superfish & CST MWS  \\ 
       \midrule
        
       Frequency (MHz) & 352.200 & 352.199    \\
      Quality Factor & 10341.6 & 10216.4  \\
     Power Dissipation (W/cm) & 123.5637 & 125.0779   \\
Kilpatrick Factor  & 1.52 & 1.53\\

       \bottomrule
   \end{tabular}
   \label{3d}
\end{table}

\section{Mechanical Construction}

\begin{figure}[htb]
   \centering
   \includegraphics*[width=29.5mm, height=25mm]{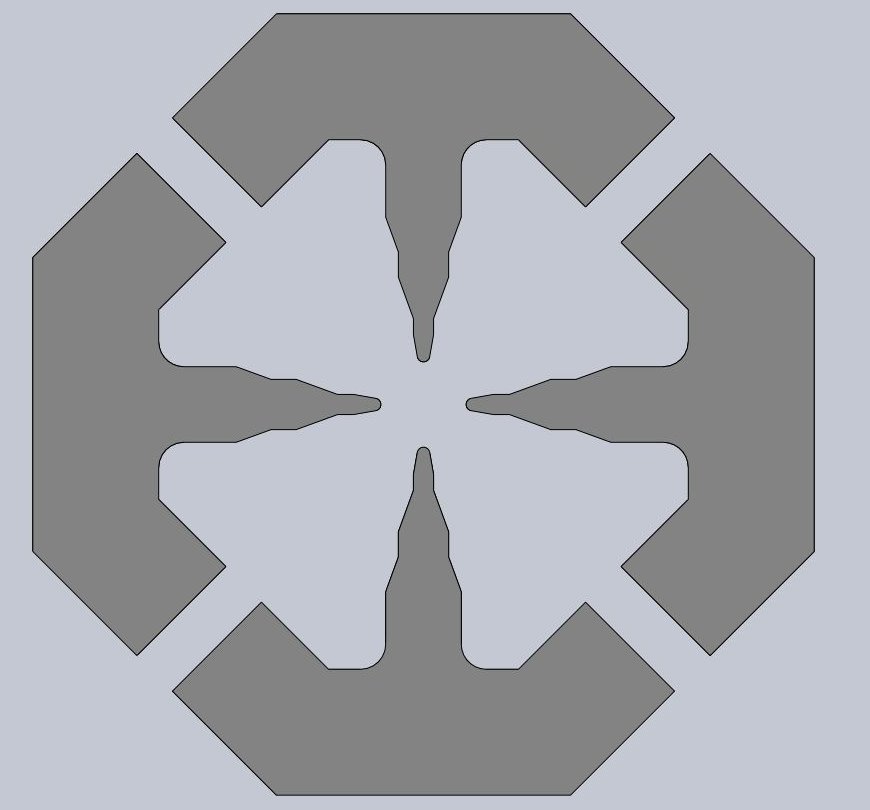}\includegraphics*[width=26mm, height=25mm]{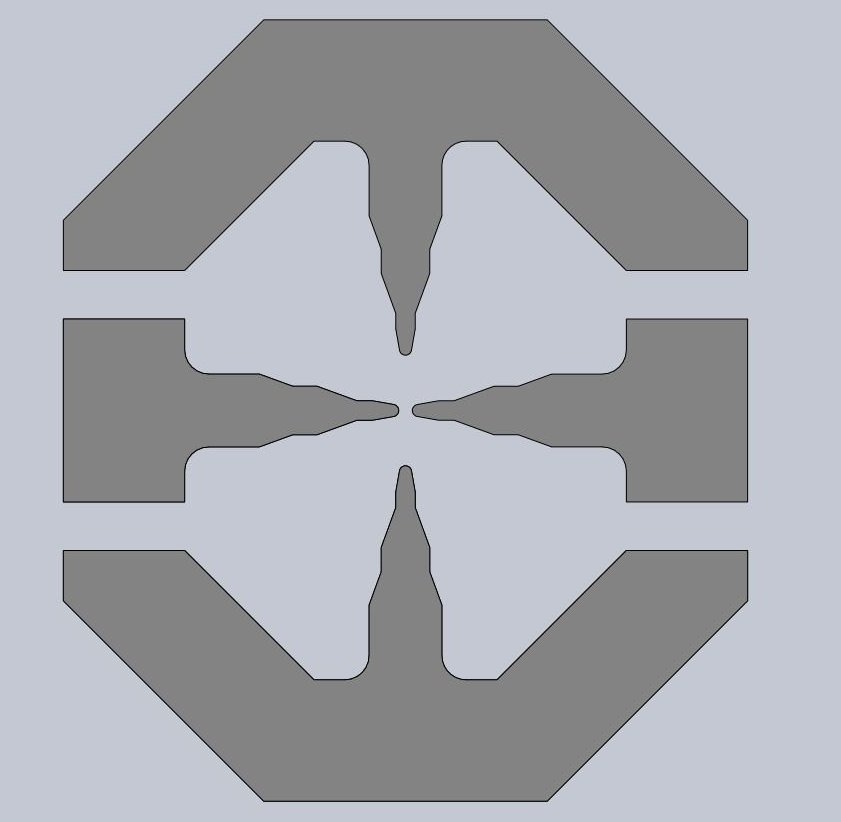}\includegraphics*[width=28mm, height=25mm]{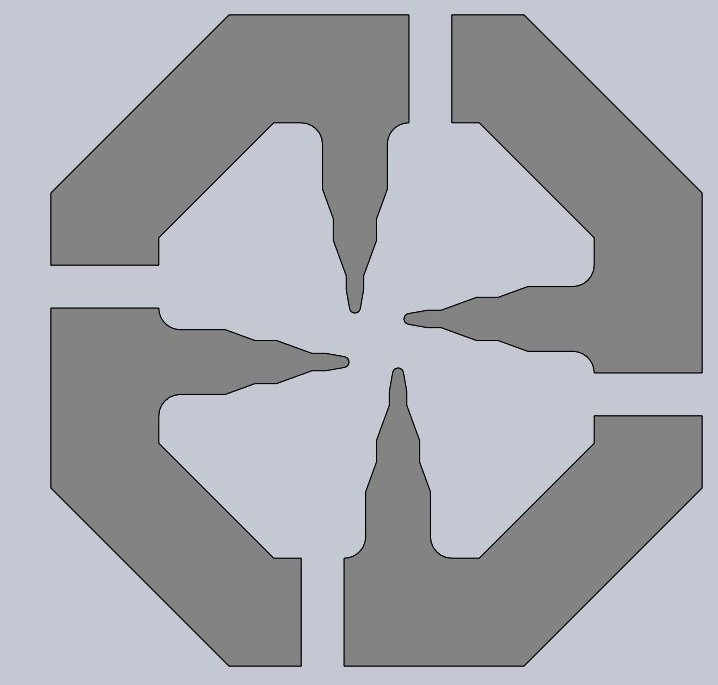}
   \caption{Different Options for Vane and Body Assembly.}
   \label{fig:Different-options-for}
\end{figure}

The choice of material, cross-sectional type and the assembly method
were driven by local manufacturing constraints and row material accessibility.
Therefore, the whole structure will be machined out of aluminum and
coated with copper except vane-tips. The decision on the quadrant
type and assembly technique has 4 criteria: CNC machining capability,
RF considerations such as contact point optimization, easiness of
assembling the manufactured pieces and laboratory work conditions.
These criteria plus the availability of enough port (vacuum, RF input,
cooling, etc...) area dictated a octagonal type geometry. Three possible
major and minor vane manufacturing options are shown in Fig. \ref{fig:Different-options-for}.
Detailed optimization and thermal analysis are ongoing to help with the
final manufacturing decision. A novel idea of using silver paste between pieces and bolting through will be used for assembling procedure. The silver paste provides high RF and
thermal conductivity and protection against high vacuum.

\section{Current Status and Conclusions}

As of September 2013, the last pieces of the ion source are being
manufactured so that it can be assembled and tested before the end
of the month. The LEBT solenoids and diagnostic stations are also
being produced with the goal of doing the basic measurements (current,
emittance, etc.) before the end of the year. As the design and production of the SANAEM RFQ is somewhat more challenging, a number of engineering models, i.e. so called cold model from Aluminum,
 will be used to understand the machining capabilities and to further test the other properties such as vacuum and low power RF. The first accelerated
protons are expected to be available by the end of 2015.

\section{ACKNOWLEDGMENT}
The authors are grateful to TAEK for their enduring support and encouragements. The authors thank A. Bozbey and A. Tanrikut for useful discussions and comments.

\end{document}